\documentclass[twocolumn,showpacs,preprintnumbers,amsmath,amssymb]{revtex4}

\usepackage{graphicx}
\usepackage{dcolumn}
\usepackage{bm}

\begin{document}

\title{Neutron Moderation in the Oklo Natural Reactor
and the Time Variation of $\alpha$\\ LA-UR 03-6686}

\author{S.K. Lamoreaux and J.R. Torgerson}

\affiliation{University of California,Los Alamos National
Laboratory,Physics Division P-23, M.S. H803, Los Alamos, NM 87545}

\date{today}

\begin{abstract}

In previous analyses of the Oklo (Gabon) natural reactor to test for a
possible time variation of the fine-structure constant $\alpha$, a
Maxwell-Boltzmann low energy neutron spectrum was assumed. We present
here an analysis where a more realistic spectrum is employed and show
that the most recent isotopic analysis of samples implies a decrease in
$\alpha$, over the last two billion years since the reactor was
operating, of $(\alpha_{past} -\alpha_{now})/\alpha \geq 4.5\times
10^{-8}$ ($6\sigma$ confidence). Issues regarding the interpretation of
the shifts of the low energy neutron absorption resonances are
discussed.

\end{abstract}
\pacs{06.20.Jr, 28.41.-i, 28.20.Gd} \maketitle

\section{Introduction}

It was recognized by Shlyakhter \cite{1} that isotopic ratios of
fission products and secondary neutron absorption reactions that
occurred with the Oklo natural reactor phenomenon some 2 billion
years ago could be used to test whether the fine-structure
constant $\alpha$ varies with time. Specifically, if a neutron
absorber (e.g., $^{149}$Sm) has a low-energy neutron absorption
resonance, and the excited compound state can decay by gamma
emission, the resonance energy will change in a predictable way if
$\alpha$ varies. The overlap of the resonance with the reactor
neutron spectrum will subsequently change thereby altering the
ratios of nearby isotopes.

In addition to \cite{1}, two other detailed analyses of the Oklo
phenomenon have been completed \cite{2,3}.  The purpose of this note is
to assess the validity of the assumption in these analyses that the
neutron spectrum can be described by a Maxwell-Boltzmann (MB)
distribution, and to determine the effects on the limit of the time
variation of $\alpha$ for different neutron spectra.  The possibility of
a non-MB spectrum was discussed in \cite{2} (footnote 3), but was not
considered to be important. However, for the high-accuracy results
presented in \cite{3}, the effects due to deviations from a MB spectrum
become very important.

\section{The Oklo Phenomenon as a Homogeneous Reactor}

The fact that the Oklo phenomenon occurred was indicated when
uranium taken from this mine was found to be deficient in
$^{235}$U; subsequent tests confirmed this discovery. From core
samples and knowledge of reactor dynamics, the age of the Oklo
phenomenon was determined to be about 2 billion years. The
conditions necessary for the multiple reactor zones to exist can
be used to place constraints upon many parameters such as
concentrations of various elements, temperature, moderator type
and concentration, {\it etc.} while the reactor was operating. For
example, it can be shown that only hydrogen or deuterium could
have served as moderators at Oklo and hydrogen present in water or
as hydration of the uranium oxide appears to be the only plausible
choice. As a first approximation to the reactor geometry, it seems
reasonable to assume that the Oklo phenomenon was a homogeneous
reactor of infinite extent. This simplifies the estimation of the
multiplication factor $k$ and the neutron flux $\Phi(E,T)$ which
we will see depends on the amount of hydrogen and neutron
absorbing impurities in the ore.

With the present isotopic ratio of $^{235}$U/$^{238}$U, the maximum $k$
that can be obtained in a homogeneous mixture of water and natural
uranium is $k=.838$, which occurs where there are 2.43 molecules of
water per atom of uranium \cite{4} (see table 18.1, p. 612). The
fractional isotopic abundance of $^{235}$U in natural U at present is
0.711\%. Extrapolating this result implies a $^{235}$U relative isotopic
concentration of about 1.2\% to achieve $k=1$. Two billion years ago,
the relative concentration of $^{235}$U was about 3.7\% (due to the
different lifetimes of $^{235}$U and $^{238}$U). This implies that two
billion years ago, $k\approx 1.3$ for a system with 2.4 H$_2$O
molecules/U atom.  If such a large fractional concentration of water was
present, the Oklo reactor would have been highly divergent.  Of the
possible effects that can reduce $k$, the two most likely are a low
water (or hydrogen) concentration in the U deposit, and the presence of
impurities. It is interesting to note that the reactor would have been
somewhat self-stabilizing as a runaway to high temperatures would drive
water out of the ore deposit.

Precise knowledge of the nature and concentration of
neutron-absorbing impurities at the time the Oklo reactor was
running is lacking. However, we can make a reasonable estimate of
the possible maximum absorption by impurities. The forgoing
arguments can made rigorous by estimating $k$ as a function of the
hydrogen atomic number density fraction, $f_H=\rho_H/\rho_U$ and
the atomic fraction of absorbing impurities $f_i=\rho_i/\rho_U$,
relative to the number density of U atoms. With $^{235}$U isotopic
fraction $x$, for thermal neutrons:
\begin{eqnarray}
k&\approx& \nu p{\rho_U x \sigma_f^{235}\over
\rho_U\left[x\sigma_a^{235}+(1-x)\sigma_a^{238}\right]+\rho_H\sigma_a^H+\sum_i
\rho_i\sigma_a^i}\nonumber\\
&\approx& 2.47 e^{-.7031 (1/f_H)^{.58}}\nonumber\\
&\times&{580x\over 690x+2.7(1-x)+0.33 f_H + \sum_i f_i\sigma_a^i}
\end{eqnarray}
where $\nu=2.47$ is the number of neutrons released per $^{235}$U
fission, $p$ is the resonance escape probability obtained from
Eqs. (10.29) and (13.3) of \cite{4}, $\sigma_f^{235}=580$ b is the
$^{235}$U fission cross section, $\sigma_a^{235}=690$ b is the
total $^{235}$U total neutron absorption cross section,
$\sigma_a^{238}=2.7$ b is the $^{238}$U total neutron absorption
cross section, $\sigma_a^H=0.33$ b is the hydrogen absorption
cross section. The total impurity absorption can be parameterized
by
\begin{equation}
\beta=\sum_i f_i\sigma_a^i
\end{equation}
Furthermore, the effects of oxygen have been neglected. For
$x=.037$, $k$ as a function of $f_H$, and as a function of
impurity concentration parameterized by $\beta$, is shown in Fig.
1. With no impurities, $k=1$ is achieved when $f_H=1.5$. For a
higher hydrogen concentration, the reactor would have run away and
the resulting high temperatures would have driven water out of the
ore; the system was likely self-stabilizing to $k=1$. As the
impurity concentration, parameterized by $\beta$ in Fig. 1,
increases, $f_H$ required for $k=1$ increases also; at $\beta=5$,
$f_H=2.6$ to achieve $k=1$. For $\beta>11$ b, it is not possible
to achieve $k\geq 1$. It is also known that the reactor operated
until $x\approx 0.02$ which implies a more restrictive limit of
$\beta \leq 3$ b, for which $f_H=5.8$.  We note that criticality
calculations at these high $^{235,238}$U concentrations are
subject to a number of corrections (see \cite{4}, pp. 463-468) so
the figures derived here are to be considered as providing
guidelines for the operation of the natural reactor which
indicates that $f_H$ is about 3, with $\beta=2$.

A ubiquitous substance that offers a significant thermal neutron
absorption cross section is sodium chloride. The thermal
absorption cross section for sodium is $\sigma_a=0.5$\,b, but for
chlorine $\sigma_a=33$\, b which limits $f_{Cl}<.1$.  The relative
atomic concentrations of other impurities can be similarly
estimated.

\section{Calculation of $\hat\sigma$ for a non-MB Neutron Spectrum}

It is well-known that when neutrons are moderated in the presence of a
$1/v$ absorber (such as U), the low-energy part of the spectrum becomes
suppressed relative to a high-energy $1/E$ component of the flux. The
relevant parameter that describes this suppression is (\cite{4}, pp.
335- 340)
\begin{equation}
\Delta={2A\Sigma_a(v_T)\over \Sigma_s}
\end{equation}
where $\Sigma_a(v_T)=\rho_a\cdot\sigma_a(v_T)$ is net absorption
cross section of the homogeneous mixture for a neutron rms
velocity corresponding to the moderator temperature,
$\Sigma_s=\rho_s\cdot\sigma_s$ is the net scattering cross section
of the moderating atoms, and $A$ is the atomic mass of the
moderating atoms. In the present case, for neutron energies about
1 eV,  the atomic mass $A=1$ for H which is responsible for most
of the neutron moderation. When $\Delta\approx 2$, the $1/E$ tail
contains the bulk of the neutrons. This is illustrated in Fig. 2
where the flux as a function of $\Delta/2$ as obtained from a
Monte Carlo calculation following the procedure in \cite{6}.

For the parameters derived in the last Section, that is $f_H=3$,
$x=0.037$, and $\beta=2$, $\Delta\approx 1$. However, for neutron
energies less than 1 eV, molecular effects become important; it is
no longer a good approximation to assume that the neutron
moderation occurs by scattering from free atoms. The chemical
bonding effect can be estimated by setting $A=2$, so in the low
neutron energy region ($.001<E<0.1$ eV), $\Delta\approx 2$ at 300
K, and scales with temperature as
\begin{equation}
\Delta=2\sqrt{300\over T}.
\end{equation}
The elements $^{149}$Sm and $^{155}$Gd have low energy neutron
resonances and are particularly interesting to test for possible time
variation of $\alpha$. The formalism developed in \cite{1,2,3} for the
effective cross section of a substance with a low-energy neutron
resonance can be understood as follows. The interaction of a neutron
with a nucleus can be parameterized through the optical model, where the
Hamiltonian is
\begin{equation}
H=U(E)+iW(E)
\end{equation}
where the absorption rate for neutrons with energy $E$ is
$\Gamma(E)=2W(E)/\hbar$. The absorption cross section can be expressed
as
\begin{equation}
\sigma(E)=\frac{g\pi}{k^2}
  \frac{\Gamma_n\Gamma_\gamma}{(E-E_r)^2/\hbar^2+{\Gamma_{tot}}^2/4}.
\end{equation}
where $\hbar k$ is the neutron momentum, $g$ is a parameter that depends
upon the spins of the neutron and interacting nuclei, $E_r$ is the
resonance energy and $\Gamma_{n,\gamma}$ are the rates for neutron and
gamma ray absorption respectively. The parameters $g$, $E_r$, and
$\Gamma_{n,\gamma}$ for elements of interest here can be found in
\cite{2,3}.

The absorption cross section at a particular energy can also be
described as the product of the rate of absorption and the times
$\tau=L/v(E)$ spent in the material, normalized to the propagation
length $L$,
\begin{equation}
\sigma(E)=\Gamma(E)/v(E).
\end{equation}
Far away from $E_r$, $\Gamma(E)$ is nearly constant and
$\sigma(E)=\sigma_0 v_0/v(E)$, where $v_0$ is the velocity at which the
cross section was measured. Nuclei whose cross sections can be described
this way are know as ``$1/v$'' absorbers. This is a more efficient
representation for thermal neutron absorption of many nuclei such as U.

The net absorption cross section is given by the integral of the cross
section time the neutron flux $\Phi(E)$,
\begin{equation}
\langle \sigma\rangle=\int \sigma(E)\Phi(E) dE.
\end{equation}
In assessing the effects of a varying low-energy resonance, because all
$1/v$ cross sections follow the same universal function which have
exactly the same form up to a constant, it is useful to define an
effective relative cross section for the nuclei with a varying
$\Gamma(E)$ as
\begin{equation}
\hat\sigma={\int \sigma(E)\Phi(E,T)dE\over \int v_0 \rho(E,T)dE}
\end{equation}
where $\sigma(E)$ is the energy-dependent cross section, $T$ is the
moderator temperature, $\Phi(E,T)$ and $\rho(E,T)$ are the neutron flux
and energy distributions, respectively, and $v_0$ is the velocity at
which $1/v$ absorption cross sections are measured.

In this analysis and the others that have been performed so far,
the finite temperature and motion of the absorbing nuclei is
ignored. This is a small correction to the resonance width because
$A>100$ for the nuclei being considered, so the absorber velocity
is much lower than the neutron velocity. Assuming $\Delta=2$ at
300 K, as a function of $T$, $\Delta=2\sqrt{300/T}$.  A procedure
similar to that described in \cite{3} (numerical integration) was
used to calculate the change in $\hat\sigma$ as a function of
change in resonance energy, but replacing the MB spectrum with
spectra derived from Fig. 2.  The same resonance parameters as
employed in \cite{3} were used, and the numerical integration was
tested with a MB spectrum for which the same results as Fig. 1 of
\cite{3} were obtained.

It is possible to determine the value of the cross section for
$^{149}$Sm, $\hat\sigma_{149}$, from detailed isotopic analyses of
core samples from Oklo and similarly for $^{155}$Gd to determine
$\hat\sigma_{155}$. Such an analysis was performed in \cite{3} and
indicates that
\begin{equation}\label{149}
\hat\sigma_{149}=91\pm 6\ {\rm kb}.
\end{equation}
This result can be used subsequently to determine the ancient value of
$E_r$ which can be compared to the present-day measured value.

Results of the numerical integration for $^{149}$Sm are shown in
Fig. 2 as a function of moderator temperature with $\Delta=2$ at
300 K. In \cite{3}, the mean temperature of the Oklo reactor was
estimated to be about 600 K.  From the curve for 600 K in Fig. 3
(expanded around the $91\pm 6$ kb in the lower plot), we find that
the change in resonance energy is
\begin{equation}\label{149result}
\Delta E_r=\left[-45\begin{array}{cc} +7\\ -15\end{array}\right]
\times 10^{-3}\ {\rm eV}
\end{equation}
which indicates more than a $6\sigma$ deviation from zero.  Note that
the``two solution" problem does not really exist in this case, but the
upper magnitude range is expanded.

In the calculation of the moderated neutron spectrum, effects of
chemical bonds (which suppress the low energy spectrum) have been
approximately included.  As the atoms become more tightly bound,
$A\rightarrow \infty$ so ($\Delta\rightarrow \infty$), and in this
limit the neutron flux spectrum becomes pure $1/E$; the result of
a calculation for this case is shown in Fig. 4. In this case, for
which two solutions exist,
\begin{equation}
\Delta E_r=(-135\pm 5)\times 10^{-3}\  {\rm eV}\ \ \ {\rm or}\ \ \
(-58\pm 5)\times 10^{-3}\ {\rm eV}.
\end{equation}
This result is roughly consistent with the $\Delta=2$ result;
reality likely is somewhere between the $\Delta=$2 and $\infty$
results. However, the result for $\Delta=2$ clearly indicates a
non-zero change in $E_r$ at a level of more than $6\sigma$.

Results for a calculation of $^{155}$Gd are shown in Fig. 5.  The
interpretation of the $^{155}$Gd data is complicated by the
presence of impurities, and an extensive analysis is given in
\cite{3}.  The Oklo data imply $\hat\sigma_{155}\approx 28\pm 10$
kb, which, from Fig. 5, indicates, for the left-hand solution,
\begin{equation}
\Delta E_r=(-90\pm 20)\times 10^{-3}\ {\rm eV}
\end{equation}
This is consistent with the $^{149}$Sm result because impurities shift
$E_r$ slightly, and the simple $1/E$ spectrum overestimates $E_r$.

\section{Interpretation of the Results}

A shift in resonance energy can be related to a shift in $\alpha$ by
\begin{equation}
{\Delta\alpha\over \alpha}={\Delta E_r\over E_0}.
\end{equation}
A reasonable estimate of $E_0=-1$ MeV was presented in \cite{2}, and is
roughly the same for $^{149}$Sm and $^{155}$Gd.  Taking the result Eq.
(\ref{149result}) determines
\begin{equation} {\alpha_{now}-\alpha_{past}\over \alpha}=
-\left[45\begin{array}{cc}+15\\-7\end{array}\right]\times
10^{-9}
\end{equation}
which is non-zero by more than $6\sigma$.  The central value plus twice
the upper uncertainty can be used to set a 95\% confidence limit on the
magnitude of the integrated rate of change over the last two billion
years
\begin{equation}
|{\dot\alpha\over\alpha}|<3.8\times 10^{-17}/{\rm yr}\ \ (\rm 95\%
conf)
\end{equation}
By assuming  higher temperatures, this limit decreases slightly.
However, a significant modification of this result requires
temperatures beyond physically reasonable values as assumed in
\cite{3}.

It is expected that the neutron absorption resonance parameters (e.g.,
the neutron and gamma widths), other than $E_r$, do not change
significantly with $\alpha$.  In estimating $E_0$ it was assumed that
$E_r$ represents a fixed energy above the ground states of nuclei with
$A$ and $A+1$, and $E_0$ is approximately the difference in the Coulomb
energies of the ground states.  This result is in agreement with a more
sophisticated analysis \cite{7}.

\section{Conclusion}

If a $1/v$ absorber is present in the moderator of a neutron chain
reactor, the neutron energy spectrum can deviate significantly
from the a Maxwell-Boltzmann distribution. The Oklo phenomenon was
rich in uranium which is a strong $1/v$ absorber. By considering
realistic deviations from a Maxwell-Boltzmann low-energy neutron
spectrum, we have shown that a recent analysis of the Oklo natural
reactor phenomenon implies an non-zero ($>6\sigma$) variation of
$\alpha$ over the last 2 billion years.  The result is of opposite
sign and an order of magnitude smaller than a recent astrophysical
determination of the change in $\alpha$ \cite{8}, which is
non-zero, but over a different time scale. It has been suggested
that the time variation of $\alpha$ is not monotonic \cite{8a}.
Our results are probably accurate to within 20\%; a full MCNP
model of the reactor, assuming reasonable estimates of the
temperature and impurities are possible, would be useful in light
of the importance and interest in the possibility of a varying
$\alpha$. Finally, these results might be interpreted more
efficiently in terms of $m_q/\Lambda_{QCD}$ for which the
sensitivity is about two orders of magnitude higher than the
sensitivity to $\alpha$ variation \cite{9}.

We thank Tuan Nguyen for critical comments and in particular for clarifying
the issues with the sign of the effect.

\begin{figure*}
\includegraphics{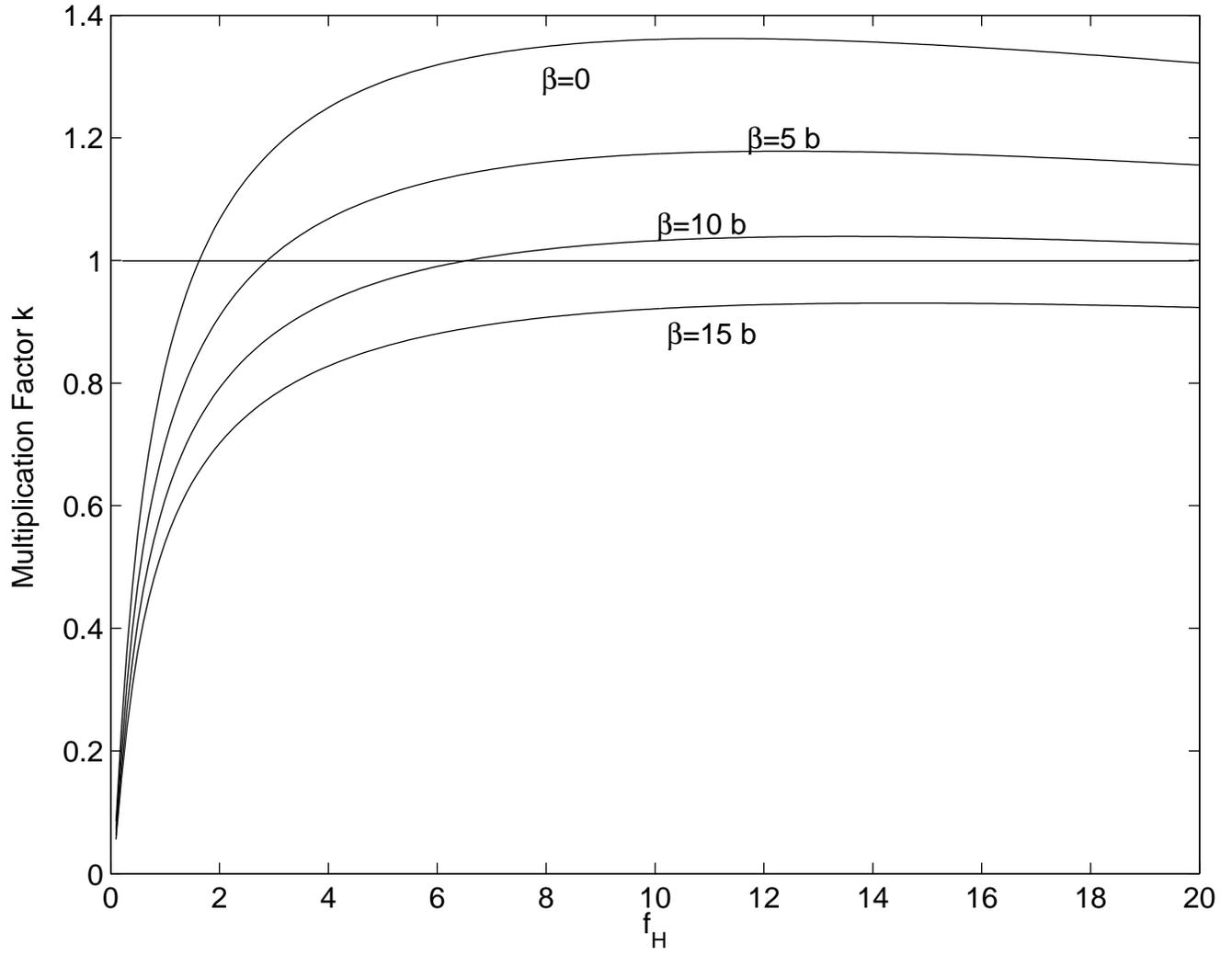}
\caption{Multiplication factor $k$ as a function of hydrogen
atomic fraction to uranium atomic fraction ($f_H$) for $^{235}$U
isotopic fraction $x=0.037$.  The plots are for increasing
$\beta=\sum_i f_i\sigma_a^i$.}
\end{figure*}

\begin{figure*}
\includegraphics{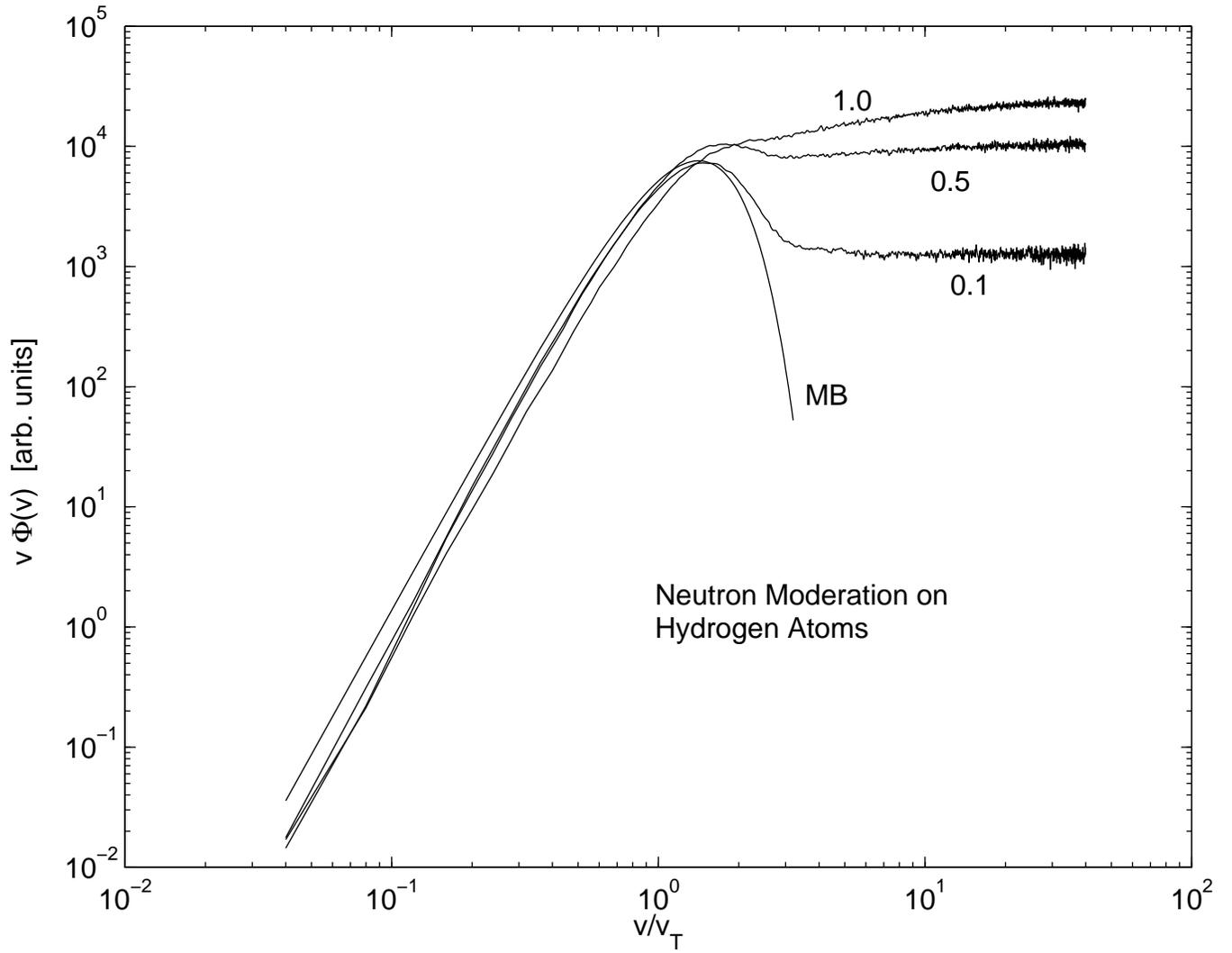}
\caption{Monte Carlo generated neutron velocity spectra for
various values of $\Delta/2=\Sigma_a(v_T)/\Sigma_s$ from 1 to .1,
where $v_T=\sqrt{2kT/m_n}$ is the rms neutron velocity, assuming
all moderation occurs from scattering on hydrogen, compared to a
MB spectrum.}
\end{figure*}

\begin{figure*}
\includegraphics{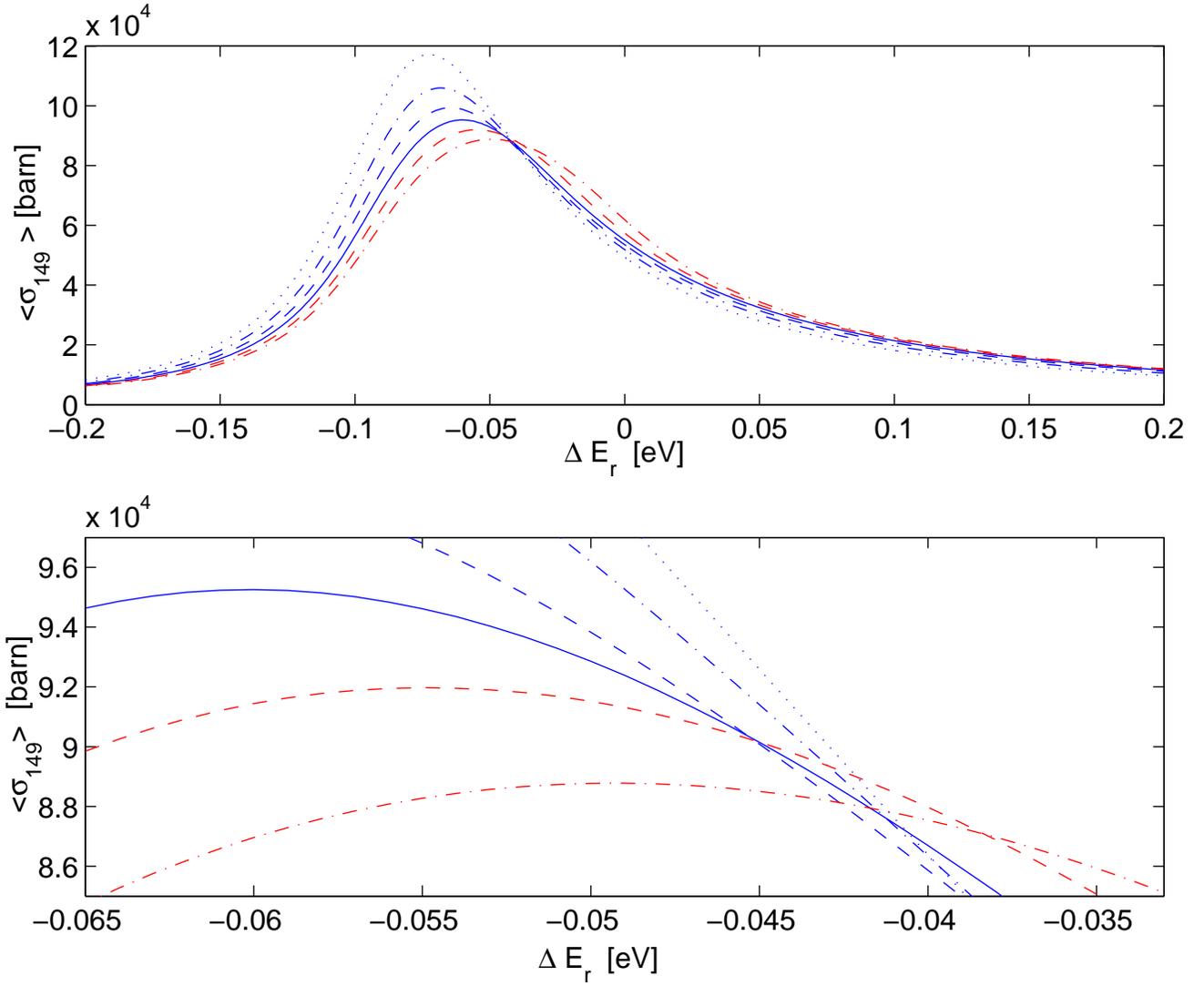}
\caption{Calculation of $\hat\sigma_{149}\equiv<\sigma_{149}>$ as
a function of change in resonance energy $E_r$ and as a function
of temperature, for $\Delta\approx 2\sqrt{300/T}$. Blue dots, 300
K; blue dot-dash: 400 K; blue dashes: 500 K; solid blue: 600 K;
red dashes: 700 K; red dot-dash: 800 K.}
\end{figure*}

\begin{figure*}
\includegraphics{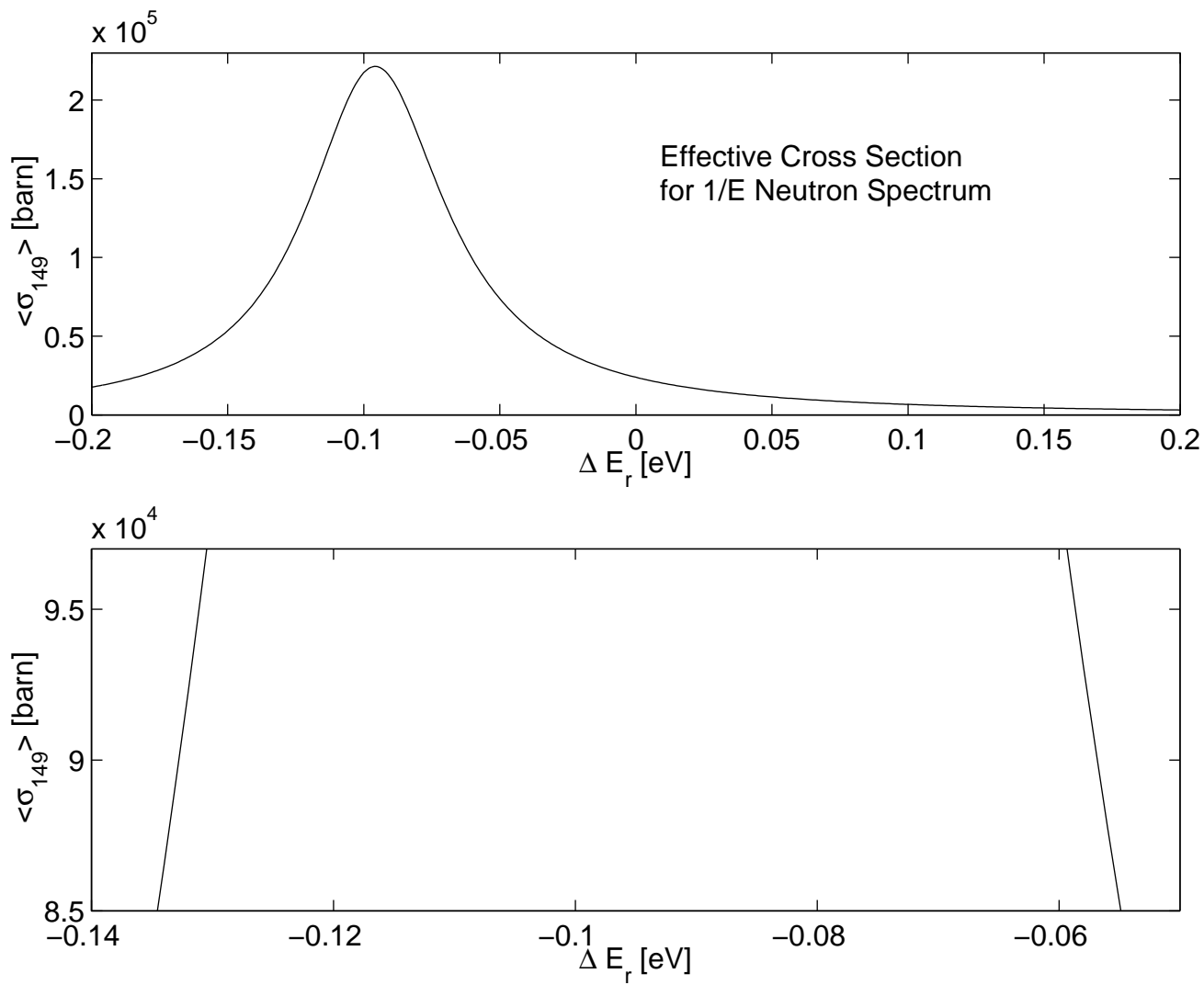}
\caption{$\hat\sigma_{149}$ assuming a $1/E$ flux spectrum.}
\end{figure*}

\begin{figure*}
\includegraphics{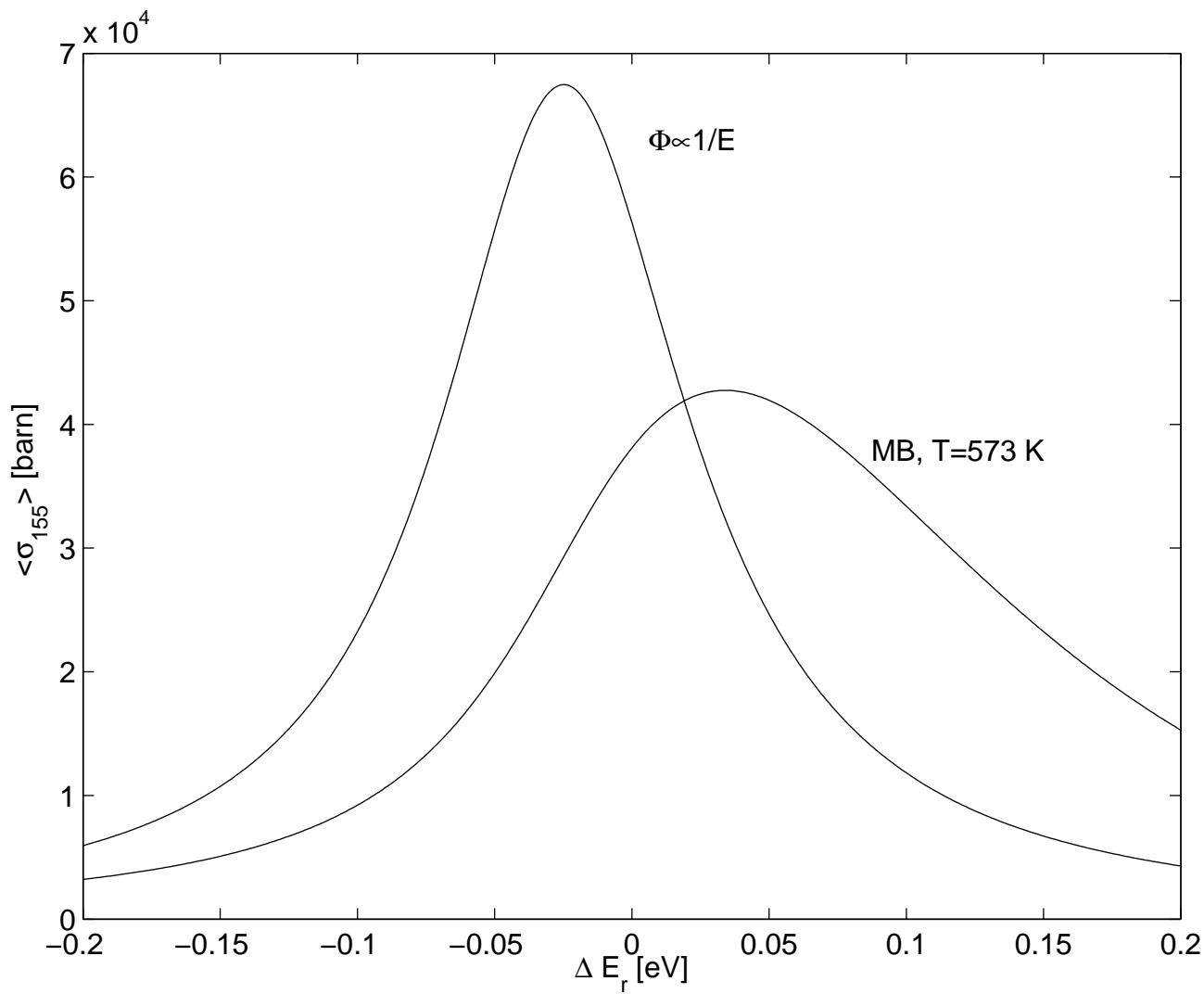}
\caption{$\hat\sigma_{155}$ for a MB spectrum ($T=$573 K) and a
$1/E$ spectrum.}
\end{figure*}

\end{document}